\documentclass[english]{lni}
\usepackage{graphicx} 

\usepackage{amsmath}
\usepackage{stmaryrd}

\usepackage{tikz}
\usepackage{todonotes}
\usetikzlibrary{positioning}
\usetikzlibrary{decorations.pathreplacing}
\usepackage{wrapfig,lipsum,booktabs}

\newcommand{\braket}[1]{\ensuremath{| #1 \rangle}}

\title{Bounds for Quantum Circuits using Logic-Based Analysis}

\begin{document}

\author[1]{Benedikt Fauseweh}{benedikt.fauseweh@tu-dortmund.de}{0000-0002-4861-7101}
\author[1]{Ben Hermann}{ben.hermann@tu-dortmund.de}{0000-0001-9848-2017}
\author[1]{Falk Howar}{falk.howar@tu-dortmund.de}{0000-0002-9524-4459}

\affil[1]{Technische Universität Dortmund}

\maketitle

\begin{abstract}
We explore ideas for scaling verification methods for quantum
circuits using SMT (Satisfiability Modulo Theories) solvers.
We propose two primary strategies: (1) decomposing proof 
obligations via compositional verification and 
(2) leveraging linear over-approximation techniques 
for gate effects. We present two examples and demonstrate 
the application of these ideas to proof Hamming 
weight preservation.
\end{abstract}


\section{Introduction}

In the formulation of quantum circuits developers struggle to design circuits that stay within their correct sub-space. 
The correct reasoning to be followed during software development is non-trivial and requires deep insight into the mechanics of the quantum program as well as their underlying theory. 
However, means to determine or proof that such a circuits stays within a certain sub-space do not exist at the moment. 

Showing that such bounds apply to a circuit would reduce the need for full quantum simulation of this circuit. 
Properties of interest here are, for instance, the Hamming distance between initial and evolved states of a circuit. 
When the Hamming distance is low, it could show us that an evolution of states over a circuit would not deviate significantly from its initial state. 
Such a property is interesting for instance for a quantum circuit description of  Many-Body Localized Discrete Time Crystals (MBL-DTCs). 
Furthermore, verification techniques can be used to show a low entanglement character of a circuit. 


We propose to determine these properties using logic-based program analysis.
Logic-based semantic descriptions of quantum circuits and verification problems based on these descriptions have been recently proposed by others, however, they lack scalability. 
They show some scalability under some optimization in the mapping of circuits to formulas (namely, so-called \emph{direct mapping}) and for simple abstractions on the Hilbert space.
We build on these ideas and establish an assume-guarantee decomposition of proof obligations to scale verification to real-world quantum programs. 
We show that these compositions hold even for non-identical but equivalent quantum circuits.
Moreover, we introduce some concrete examples of properties.

The novel contribution of our work is the formulation of the analysis or verification problem in a compositional manner. 
We decompose the proof obligations and show that we can establish compositional bounds for (partial) circuits. 
We approach the inherent scalability problem of quantum program analysis with this compositionality as well as linear over-approximation of gate effects. 
Such an over-approximation enables us to provide meaningful bound in reasonable time. 


\section{Related Work}
The derivation of properties for quantum programs or circuits has been studied before. We give an overview over current verification techniques and their applications. Proofs of these properties can be massively supported by judgements on circuit equivalence and symmetry. 

\paragraph*{Verification Techniques}

Bauer-Marquart et al.~\cite{10.1007/978-3-031-27481-7_12} provide a framework for symbolic verification of quantum programs. Verification problems can be expressed as an SMT formula checked by a $\delta$-complete solver in their approach. We draw inspiration from their work and also express circuits in such a manner, but apply a linear over-approximation in order to scale better. 
Takagi et al.~\cite{10.1007/978-3-031-51777-8_5} extend Dynamic Quantum Logic in order to automate verification. They use bra-ket notation instead of complex vectors and matrices, thus, relying on a more compact representation. 
Chareton et al.~\cite{10.1007/978-3-030-72019-3_6} contribute QBricks, a first-order-logic-based framework for the specification and proof of quantum programs. They also use a symbolic representation of quantum states.
Feng and Xu~\cite{10.1145/3582016.3582039} present a verification system for non-deterministic quantu programs based on Hoare-style logic. 
Lewis et al.~\cite{10.1145/3624483} provide a survey on verification techniques for quantum programs. Sarkar~\cite{sarkar_automated_2024} outlines a research vision for automated quantum software engineering where verification of quantum programs is a key step in the support of developers. 




\paragraph*{Equivalence and Symmetry}

Verification tasks for quantum circuits can be aided with a proof of circuit equivalence. 
Mei et al.~\cite{10.1007/978-3-031-63501-4_21} provide a reduction of the (universal) quantum circuits equivalence problem to weighted model counting (WMC). This reduction outperforms classic equivalence checking based on the ZX calculus. 
Anselmetti et al.~\cite{Anselmetti_2021} propose gate fabrics the express equivalent behavior as complex circuits for the the simulation of strongly correlated ground and excited states of molecules and materials under the Jordan–Wigner mapping. The gate fabrics can be implemented linearly locally and preserve all relevant quantum numbers. 

Another helpful property is symmetry preservation. 
Gard et al.~\cite{Gard2020} present general schemes to facilitate state preservation circuits for quantum simulation that respect a number of symmetries. This effectively reduces the Hilbert space to explore in simulations. 









\section{Motivating Example}

\begin{figure}[t]
    \centering
      \includegraphics[width=1.00\columnwidth]{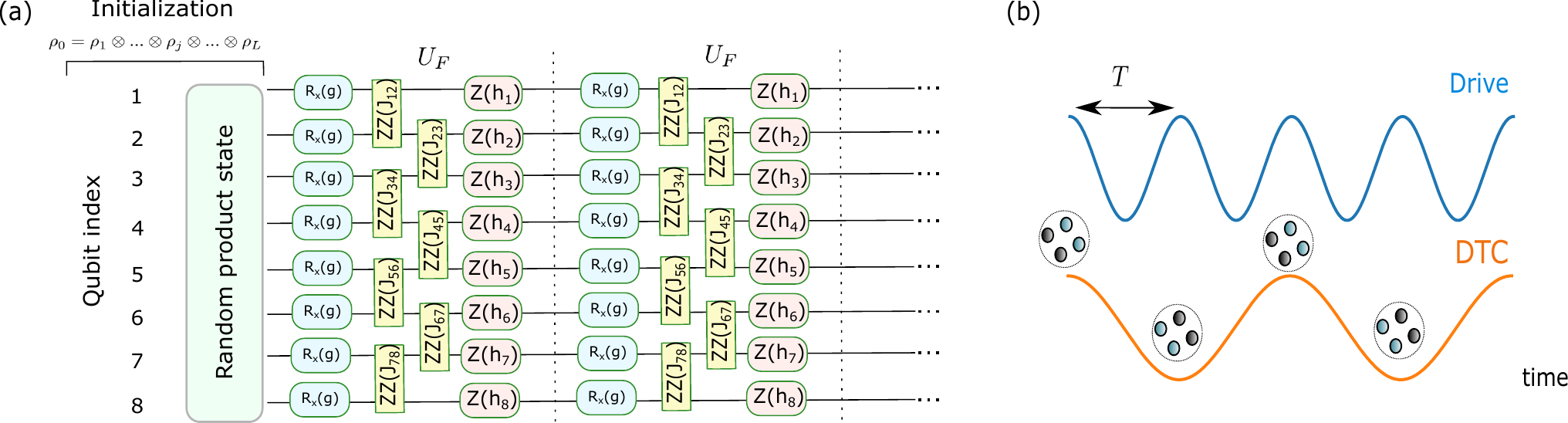}
      \caption[]
			{ (a) Quantum circuit for the kicked Ising MBL-DTC. (b) A time crystals spontaneously breaks the time translational invariance of the drive. }
      \label{f:circuit_mbl_dct}
    \end{figure}
    
The study of Many-Body Localized Discrete Time Crystals (MBL-DTCs) provides a motivating example for our approach on higher-level programming abstractions in quantum software engineering \cite{Basermann2024}. MBL-DTCs represent a non-equilibrium phase of matter characterized by a spontaneous breaking of discrete time-translation symmetry, resulting in a subharmonic response that spontaneously breaks the periodicity of an external drive \cite{khemani2019briefhistorytimecrystals}. This phenomenon emerges from the interplay between many-body localization (MBL), the generalization of Anderson localization \cite{RevModPhys.80.1355} to interacting systems, and external periodic driving, leading to a form of spatiotemporal order in quantum systems.

Despite significant interest, the existence of MBL-DTCs remains an open question due to the potential instability of the underlying MBL phase. Specifically, MBL systems may be susceptible to an “avalanche” mechanism, where rare regions of ergodicity can lead to thermalization over very long timescales \cite{PhysRevB.95.155129,PhysRevB.105.174205,PhysRevB.106.L020202,PhysRevB.108.134204}. Traditional analytical proofs addressing this issue are complex and challenging to verify, while numerical simulations are limited to small system sizes that cannot capture the avalanche mechanism in the thermodynamic limit.

MBL-DTCs can be effectively described using quantum circuits, making them amenable to abstract analysis techniques. Recent work has demonstrated that quantum computers can be programmed to realize the DTC phase \cite{PRXQuantum.2.030346,PhysRevResearch.6.033092,Fauseweh2024} and experimentally detect its dynamical properties \cite{Mi2022}, leveraging extensive capabilities in programmability, initialization, and readout. Additionally, evidence for MBL-DTCs in higher dimensions have been presented using infinite tensor network states \cite{PhysRevB.103.224205}.

As a typical example for an MBL-DTC we consider the one dimensional periodically kicked Ising model,
\begin{eqnarray}\label{eq:floquet_op}
U_F=e^{-i \frac{T}{4}\sum_{j=1}^{L-1} J_j\sigma_j^z\sigma_{j+1}^{z}}e^{-i \frac{T}{2}\sum_{j=1}^{L}h_j\sigma^{z}_j}e^{-i \frac{\pi g}{2}T\sum_{j=1}^L\sigma^{x}_{j}}.
\end{eqnarray}
This unitary directly translates to a quantum circuit on a 1D topology of qubits. The parameter $T$ represents the Floquet period of the external drive, which we set to $T=1$. The parameter $g$ is the pulse parameter, with $g=1$ representing a perfect bit flip. The coupling parameters $J_j$ and the magnetic fields $h_j$ are randomly sampled from uniform distributions \cite{Mi2022} with $J_j\in\left[-1.5\pi,-0.5\pi\right]$ and $h_j\in[-\pi,\pi]$.

Applying abstract analysis methods to these quantum circuits can be used to verify preserved properties. By focusing on specific quantities that characterize the MBL-DTC phase, we can perform logic-based analyses without the need for full quantum simulations \cite{Fauseweh2021}, making only statements about limited but insightful aspects of the system’s behavior.

One such property is the Hamming distance between the initial and evolved states. The Hamming distance  $d$  quantifies the minimum number of bit flips required to transform one bitstring into another. In MBL-DTC systems, unlike in ergodic dynamics where an initial bitstring state quickly spreads over the entire computational basis, the evolution prevents the state from deviating significantly from its initial configuration. 

Another critical property is the low entanglement characteristic of the MBL-DTC phase. Given that states in this phase do not significantly deviate from their initial values, the entanglement entropy of initially random bit string states remains low compared to that in ergodic phases, where entanglement rapidly increases. This property can be analyzed using reduced density matrices of subsystems. It has been shown that abstract interpretation can effectively analyse these properties in a scalable way \cite{10.1145/3453483.3454061}. By proving that the entanglement entropy remains low, we can infer that the system retains localization properties over time. 

Hence applying logic based analysis to MBL-DTCs could provide valuable insights into the conditions necessary for their stability against thermalization and avalanches. This methodology allows for the decomposition of complex verification tasks into manageable sub-tasks, enabling the analysis of large-scale quantum circuits that model MBL-DTCs.


\section{Preliminaries}

We briefly introduce some basic formalization of computations
on qubits and logic-based verification techniques.

\subsection{Qubits and Quantum Circuits}

\paragraph*{Qubits}
A qubit, the fundamental unit of quantum information, is represented as a vector in a two-dimensional complex Hilbert space. Using Dirac’s bra–ket notation, the state of a qubit $\braket{q}$ can be expressed as a superposition of the computational basis states $\braket{0}$ and $\braket{1}$, $\braket{q} = \alpha\braket{0} + \beta\braket{1},$ where $\alpha, \beta \in \mathbb{C}$ are complex coefficients satisfying the normalization condition $|\alpha|^2 + |\beta|^2 = 1$. The basis states $\braket{0}$ and $\braket{1}$ form an orthonormal basis of the Hilbert space $\mathbb{C}^2$, allowing for the full description of any qubit state within this space.
For a system of $n>1$ qubits, we can compute the state 
from the coefficients of the individual qubits as
the nested Kronecker product
$\bigotimes_{1\leq i \leq n} \braket{q_i}$. 
This representation, however, does not lend itself to 
analysis with constraint solvers as the 
repeated multiplication of coefficients results 
in many non-linear expressions.
We can alternatively represent the state of
the system by $2^n$ complex coefficients of its $2^n$
basis states. We iterate with $c_{b(i)}$ over these 
coefficients, where $b(i)= i_{\mathbf{2}}$ for $0\leq i< 2^n$.
For a two-qubit system, e.g., the four basis states are 
$\braket{00}$, $\braket{01}$, $\braket{10}$, and 
$\braket{11}$.
While this representation avoids non-linear expressions
in the state, it requires exponentially many variables
(in the number of qubits) and will likely not 
scale to big quantum circuits.


\paragraph*{Quantum Circuits}
Quantum circuits apply so-called \emph{gates} to individual
qubits or two pairs of qubits. Some commonly used gates are 
the Hadamard gate, rotations, or Pauli gates. We express the 
effect of gates on qubits as unitary matrices.
The Hadamard gate $H$, e.g. is a single-qubit operation that 
maps the basis state $\braket{0}$ to $\frac{\braket{0}+\braket{1}}{\sqrt{2}}$
and $\braket{1}$ to $\frac{\braket{0}-\braket{1}}{\sqrt{2}}$
creating an equal superposition of the two basis states.
Other gates rotate qubits around some axis and are parameterized 
by the angle of rotations, e.g. $R_y(\theta)$ rotates a qubit 
around the $y$-axis by $\theta$ and can be represented by the 
matrix.
Gates that entangle necessarily work on multiple qubits, i.e.,
at least on two qubits. The controlled Z (CZ) gate for example
uses one qubit as the control and changes the relative phase 
of the target qubit when the control qubit is in the state 
$\braket{1}$.
We compute the effect of gates on states as matrix-vector products
of gates and states.
Quantum circuits will typically also contain measurements which
collapse superpositions and read out classical results of 
computations. 

\paragraph*{Sub-Space Preservation}
We are interested in showing that a quantum circuit 
preserves the sub-space of the input. 
These properties are 
relevant in many applications, e.g. sub-spaces determined by symmetry in digital quantum simulation \cite{lively2024robustexperimentalsignaturesphase, PRXQuantum.5.020328} or constrained sub-spaces in combinatorial optimization \cite{a15060202}.
One simple example is the preservation of the Hamming weight of a quantum state.
The expected Hamming weight of a quantum state
is the weighted sum of its basis states.
Consider a general \( n \)-qubit state
$\braket{\psi} = \sum_{i=0}^{2^n-1} c_i \braket{i}$
where $c_i$ are complex coefficients, and $\braket{i}$ represents the computational basis state corresponding to the binary representation of $i$. The probability of each basis state is $|c_i|^2$.
For each computational basis state $\braket{i}$, the Hamming weight $w(i)$ is the number of qubits in the state $\braket{1}$. The expected Hamming weight of the quantum state $\braket{\psi}$ can be calculated as:
   \[
\mathrm{HW}(~\braket{\psi}~)= \sum_{i=0}^{2^n-1} w(i) \cdot |c_i|^2
   \]
For a two-qubit state $\braket{\psi} = c_0 \braket{00} + c_1 \braket{01} + c_2 \braket{10}+ c_3 \braket{11}$, 
the expected Hamming weight is
$0 \cdot |c_0|^2 + 1 \cdot |c_1|^2 + 1 \cdot |c_2|^2 + 2 \cdot |c_3|^2
$.
There, the complex absolute square $|a+ib|^2$ can be expressed by the quadratic expression $a^2+b^2$.

\subsection{Logic-based Verification of Quantum Circuits}

\paragraph*{First Order Logic and Theories}

For a set $\mathbf{x}$ of variables, let 
$\varphi[\mathbf{x}]$ be a logic formula in some theory 
$T$ over variables $\mathbf{x}$. The
theory provides a signature $\Sigma$ defining constants, functions, and relations, as well as a set of axioms $A_T$ (i.e., closed formulae) that constrain the interpretation of the signature. 
A model $M = (D, I)$ is a pair
of domain $D$ and interpretation $I$.
There $I$ maps elements of the signature to
concrete constants, functions, and relations
on $D$.
We call $v:\mathbf{x} \rightarrow D$,
mapping variables to values in $D$, a variable assignment.
A model $M$ is then a $T$-model, if $\llbracket\varphi\rrbracket_{M,v}$ is true for all $\varphi\in A_T$ and assignments $v$,
where $\llbracket\cdot\rrbracket_{M,v}$ denotes the 
evaluation under $M$ and $v$.
As an example, the formula $(x=0)$ is true for
$v(x)=0$ and
model $M$ with $D=\mathbb{R}$ and $I(=)$ the usual
interpretation of \emph{equality} of real numbers.
A $\Sigma$-formula $\varphi$ is $T$-satisfiable,
if there is a $T$-model $M$ for which $\llbracket\varphi\rrbracket_{M}$ is true ---
the formula is $T$-valid if this is the case for every model $M$.
We say that a formula $\varphi$ entails another formula $\psi$,
denoted by $\varphi \models \psi$ if 
$\psi$ is true when $\varphi$ is true,
i.e., when the formula $(\neg \varphi \lor \psi)$ is $T$-valid or equivalently $(\varphi \land \neg \psi)$ is
not $T$-satisfiable. SMT solvers implement decision
procedures for deciding satisfiability for certain theories.
Two theories that we use in this work are $QF\_LRA$, 
the quantifier-free fragment of linear real arithmetic
and $QF\_NRA$, the quantifier-free fragment of the non-linear 
real arithmetic. Some SMT solvers offer approximating 
analysis for extensions of these theories, e.g., with
support for trigonometric functions.

% ...

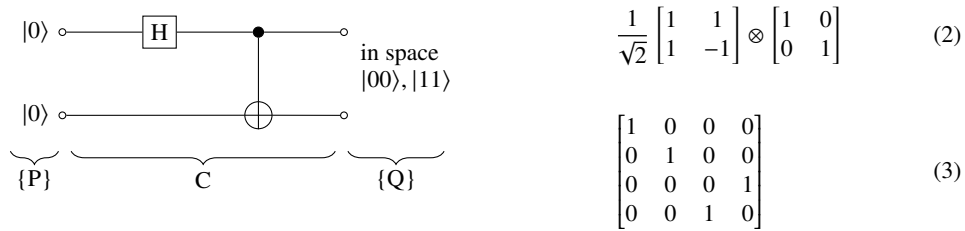
\begin{figure}[t!]
\centering
\begin{minipage}{.45\textwidth}
\begin{tikzpicture}

\node[draw,circle,inner sep=0.1em] (q1) {};
\node[draw,circle,inner sep=0.1em,below=of q1] (q2) {};

\node[draw,rectangle,right=of q1] (h) {H};

\node[draw,circle,inner sep=0.15em,fill=black,right=of h] (cnot1) {};

\node[draw,circle,inner sep=0.4em,below=of cnot1,yshift=.5em] (cnot2) {};

\node[draw,circle,inner sep=0.1em,right=of cnot1] (o1) {};
\node[draw,circle,inner sep=0.1em,below=of o1] (o2) {};

\draw (q1) -- (h.west);
\draw (h.east) -- (cnot1.west);
\draw (q2) -- (cnot2.west);

\draw (cnot1.east) -- (o1.west);
\draw (cnot2.east) -- (o2.west);

\draw (cnot1.south) -- (cnot2.north);
\draw (cnot2.south) -- (cnot2.north);
\draw (cnot2.west) -- (cnot2.east);

\node[left] at (q1.west) {$\braket{0}$};
\node[left] at (q2.west) {$\braket{0}$};

\node[right,align=left] at ([xshift=.5em,yshift=-1.5em]o1.west) {in space\\ $\braket{00}, \braket{11}$};

\draw [decorate,decoration={brace,amplitude=5pt,mirror,raise=4ex}]
  ([xshift=-2em,yshift=.3em]q2.west) -- ([yshift=.3em]q2.west) node[midway,yshift=-3em]{\{P\}};

\draw [decorate,decoration={brace,amplitude=5pt,mirror,raise=4ex}]
  ([yshift=.3em]o2.east) -- ([xshift=4em,yshift=.3em]o2.east) node[midway,yshift=-3em]{\{Q\}};

\draw [decorate,decoration={brace,amplitude=5pt,mirror,raise=4ex}]
  ([xshift=.2em,yshift=.3em]q2.east) -- ([xshift=-.2em,yshift=.3em]o2.west) node[midway,yshift=-3em]{C};


\end{tikzpicture}
\end{minipage}
\hfill
\begin{minipage}{.4\textwidth}
\begin{align}
\frac{1}{\sqrt{2}}
\begin{bmatrix}
1 & 1\\
1 & -1\\
\end{bmatrix}
\otimes
\begin{bmatrix}
1 & 0 \\
0 & 1 \\
\end{bmatrix}
\end{align}
\begin{align}
\begin{bmatrix}
1 & 0 & 0 & 0 \\
0 & 1 & 0 & 0 \\
0 & 0 & 0 & 1 \\
0 & 0 & 1 & 0 \\
\end{bmatrix}
\end{align}
\end{minipage}
\caption{Left: two-qubit curcuit $C$ with pre-condition $P$ and post-condition $Q$. Right: Matrix representations of 
Hadamard gate (1) and CNOT gate (2).}
\label{fig:example1}
\end{figure}

\paragraph*{Logic-based Verification}

When analysing quantum circuits, we are interested in 
showing that, given some condition $P$ on the on the 
inputs of the circuit, some condition $Q$ holds
on the outputs. 
We refer to $P$ as the precondition and to $Q$
as the post-condition. The verification problem is
then to decide if $\{P\}C\{Q\}$ is a (so-called) 
valid Hoare-triple for the circuit $C$.
To this end, we encode the circuit and the conditions as
logic formulas and then decide the entailment 
$(P\land C) \models Q$, which holds if 
$(P\land C) \land \neg Q$ is not satisfiable.
We do not define the translation formally here but 
only provide some intuition. We refer
readers to previous work by others~\cite{10.1007/978-3-031-27481-7_12} for a detailed formal account.

In the example shown in 
Figure~\ref{fig:example1} we are interested in proving 
that the state after the circuit is in the subspace
spanned by $\braket{00}$ and $\braket{11}$ if 
the two qubits are $\braket{0}$ initially.
Figure~\ref{fig:constraints} shows the logic encoding for this small two-qubit circuit.
We encode a quantum circuit in variables and constraints
by introducing complex variables for the coefficients $c_i$
of the basis states at different stages of the computation,
with $c_i^0$ representing the initial value of $c_i$
over which we express the pre-condition and $c_i^k$ 
represents the value after $k$ steps. For $0< j \leq k$,
we encode the effect of step $j$ of the circuit (i.e.,
the effect of some gate) as a logic formula over 
variables $c_i^{j-1}$ and $c_i^{j}$. 




\begin{figure}
\begin{align*}
P  := ~ & (c_{00}^{0} = 1) ~ \land~  (c_{01}^{0} = 0)
~ \land ~ (c_{10}^{0} = 0) ~ \land~  (c_{11}^{0} = 0) \\
C  := ~  & (c_{00}^{2} =  c_{00}^{1})
~ \land ~
(c_{00}^{1} = \frac{1}{\sqrt{2}}(c_{00}^{0} + c_{10}^{0})) ~ \land 
  (c_{01}^{2} =  c_{01}^{1})
~ \land ~
(c_{01}^{1} = \frac{1}{\sqrt{2}}(c_{01}^{0} + c_{11}^{0})) ~ \land ~ \\
& (c_{10}^{2} =  c_{11}^{1})
~ \land ~
(c_{10}^{1} = \frac{1}{\sqrt{2}}(c_{00}^{0} - c_{10}^{0})) ~ 
~ \land ~ 
  (c_{11}^{2} =  c_{10}^{1})
~ \land ~ 
(c_{11}^{1} = \frac{1}{\sqrt{2}}(c_{01}^{0} - c_{11}^{0}))  \\
Q  := ~ & (c_{01}^{2} = 0) ~ \land~  (c_{10}^{2} = 0)
\end{align*}

\caption{Logic encoding of the quantum circuit C in Figure~\ref{fig:example1} along with pre-condition
P and post-condition Q, 
expressed as first-order formulas over the 
complex coefficients of the basis states.
This representation is 
translated to formulas over reals to decide 
$P\land C \models Q$ with an SMT solver.}
\label{fig:constraints}
\end{figure}








\section{Compositional Verification of Quantum Circuits}

The SMT-based verification of quantum circuits 
presented in the previous section will not scale 
to bigger circuits due to large numbers of variables
and constraints that are required to encode these
circuits. 
We propose two strategies for scaling verification: (1) decomposition of proof obligations
and (2) over-approximation of gate
effects, drawing inspiration from the application
of these techniques for the verification of 
airplane control systems~\cite{DBLP:conf/fm/BratBDGHK15}.

\begin{figure}[t!]
\footnotesize
\begin{align*}
P := ~ & (c_{00}^{0} = 1) ~ \land~  (c_{01}^{0} = 0) ~ \land ~
  (c_{10}^{0} = 0) ~ \land~  (c_{11}^{0} = 0) \\
A_1  := ~ &  (\frac{1}{\sqrt{2}}(c_{01}^{0} + c_{11}^{0}) = 0) ~ \land~ (\frac{1}{\sqrt{2}}(c_{01}^{0} - c_{11}^{0}) = 0) \\
C_1  := ~  & 
(c_{00}^{1} = \frac{1}{\sqrt{2}}(c_{00}^{0} + c_{10}^{0}))  \land 
(c_{01}^{1} = \frac{1}{\sqrt{2}}(c_{01}^{0} + c_{11}^{0}))  \land 
(c_{10}^{1} = \frac{1}{\sqrt{2}}(c_{00}^{0} - c_{10}^{0})) 
 \land 
(c_{11}^{1} = \frac{1}{\sqrt{2}}(c_{01}^{0} - c_{11}^{0}))  \\
%
A_2  := ~ & (c_{01}^{1} = 0)  \land  (c_{11}^{1} = 0) \\
%
C_2 := ~  & 
(c_{00}^{2} =  c_{00}^{1})
 \land   (c_{01}^{2} =  c_{01}^{1})
 \land 
 (c_{10}^{2} =  c_{11}^{1})
 \land  (c_{11}^{2} =  c_{10}^{1})
\\
Q  := ~ & (c_{01}^{2} = 0)  \land  (c_{10}^{2} = 0)
\\
 \mbox{ }
 \\
& \mbox{Compositional proofs:} \qquad P \models A_1  \qquad 
A_1 \land C_1 \models A_2  \qquad 
A_2 \land CNOT \models Q
\end{align*}

\caption{Decomposition of the proof from Figure~\ref{fig:constraints}.}
\label{fig:decomposed}
\end{figure}

\paragraph*{Compositional Verification}

In compositional verification 
we decompose a circuit
$C$ into sub-circuits $C_1, \ldots, C_n$,
such that $C=C_1;\ldots;C_2$,  i.e., $C$ is the 
sequential composition of the sub-circuits. We then
introduce local properties $A_1, \ldots, A_n$ such 
that $P \models A_1$, and $A_i \land C_i \models A_{i+1}$
for $1 \leq i < n$, and $A_n \land C_n \models Q$.
In the example from the previous section, 
we could decompose the proof as shown in Figure~\ref{fig:decomposed}.
into three parts.
Proving properties in this style 
will be especially scalable if (a) component-level
proofs do not have to be computed on the complete state but only on the affected qubits and if
(b) appropriate intermediate guarantees can be computed in an automated way.

We conjecture that 
for many gates, assumptions can be computed using the
gate matrix and the post-condition in style of a weakest 
precondition predicate transformer. In fact, the assumptions in the  example are exactly the weakest pre-conditions that are obtained by replacing equal terms. 
In these
cases, we would not even have to use an SMT solver to proof an obligation as the assumption results from a syntactic transformation and is the weakest pre-condition by 
construction. 
%
%
%

\begin{wrapfigure}{r}{4cm}
\begin{align*}
 H(4) = \begin{bmatrix}
1 & 0 & 0 & 0 \\
0 & c & +s & 0 \\
0 & -s & c & 0 \\
0 & 0 & 0 & 1 \\
\end{bmatrix} \\
\textbf{}\mbox{for} \quad
\begin{matrix}
c ~:=~\cos(\lambda/2)\\
s ~:=~\sin(\lambda/2)\\
\end{matrix}
\end{align*}
\caption{H(4) gate matrix for parameter $\lambda$.}
\label{fig:approx}
\end{wrapfigure}

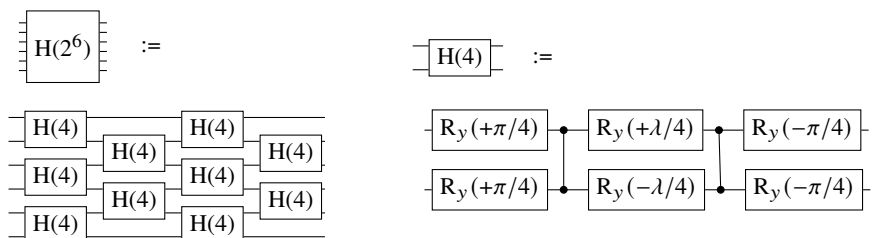
\begin{figure}[t!]
\begin{tikzpicture}

\node[inner sep=0.1em] (q1) {};
\node[inner sep=0.1em] (q2) at ([yshift=-1em]q1){};
\node[inner sep=0.1em] (q3) at ([yshift=-1em]q2){};
\node[inner sep=0.1em] (q4) at ([yshift=-1em]q3){};
\node[inner sep=0.1em] (q5) at ([yshift=-1em]q4){};
\node[inner sep=0.1em] (q6) at ([yshift=-1em]q5){};

\node[inner sep=0.1em, right=of q1,xshift=10em] (o1) {};
\node[inner sep=0.1em] (o2) at ([yshift=-1em]o1){};
\node[inner sep=0.1em] (o3) at ([yshift=-1em]o2){};
\node[inner sep=0.1em] (o4) at ([yshift=-1em]o3){};
\node[inner sep=0.1em] (o5) at ([yshift=-1em]o4){};
\node[inner sep=0.1em] (o6) at ([yshift=-1em]o5){};

\node[draw,rectangle,right=of q1,yshift=-.5em,xshift=-2.5em] (h1) {H(4)};
\node[draw,rectangle,right=of q3,yshift=-.5em,xshift=-2.5em] (h2) {H(4)};
\node[draw,rectangle,right=of q5,yshift=-.5em,xshift=-2.5em] (h3) {H(4)};

\node[draw,rectangle,right=of h1,yshift=-1em,xshift=-2.5em] (h4) {H(4)};
\node[draw,rectangle,right=of h2,yshift=-1em,xshift=-2.5em] (h5) {H(4)};

\node[draw,rectangle,right=of h4,yshift=1em,xshift=-2.5em] (h6) {H(4)};
\node[draw,rectangle,right=of h5,yshift=1em,xshift=-2.5em] (h7) {H(4)};
\node[draw,rectangle,right=of h5,yshift=-1em,xshift=-2.5em] (h8) {H(4)};

\node[draw,rectangle,right=of h6,yshift=-1em,xshift=-2.5em] (h9) {H(4)};
\node[draw,rectangle,right=of h7,yshift=-1em,xshift=-2.5em] (h10) {H(4)};

\draw (q1) -- ([yshift=.5em]h1.west);
\draw (q2) -- ([yshift=-.5em]h1.west);

\draw (q3) -- ([yshift=.5em]h2.west);
\draw (q4) -- ([yshift=-.5em]h2.west);

\draw (q5) -- ([yshift=.5em]h3.west);
\draw (q6) -- ([yshift=-.5em]h3.west);

\draw ([yshift=.5em]h1.east) -- ([yshift=.5em]h6.west);
\draw ([yshift=-.5em]h1.east) -- ([yshift=.5em]h4.west);
\draw ([yshift=.5em]h2.east) -- ([yshift=-.5em]h4.west);
\draw ([yshift=-.5em]h2.east) -- ([yshift=.5em]h5.west);
\draw ([yshift=.5em]h3.east) -- ([yshift=-.5em]h5.west);
\draw ([yshift=-.5em]h3.east) -- ([yshift=-.5em]h8.west);

\draw ([yshift=.5em]h6.east) -- (o1);
\draw ([yshift=-.5em]h6.east) -- ([yshift=.5em]h9.west);
\draw ([yshift=.5em]h7.east) -- ([yshift=-.5em]h9.west);
\draw ([yshift=-.5em]h7.east) -- ([yshift=.5em]h10.west);
\draw ([yshift=.5em]h8.east) -- ([yshift=-.5em]h10.west);
\draw ([yshift=-.5em]h8.east) -- (o6);

\draw ([yshift=.5em]h4.east) -- ([yshift=-.5em]h6.west);
\draw ([yshift=-.5em]h4.east) -- ([yshift=.5em]h7.west);
\draw ([yshift=.5em]h5.east) -- ([yshift=-.5em]h7.west);
\draw ([yshift=-.5em]h5.east) -- ([yshift=.5em]h8.west);

\draw ([yshift=.5em]h9.east) -- (o2);
\draw ([yshift=-.5em]h9.east) -- (o3);
\draw ([yshift=.5em]h10.east) -- (o4);
\draw ([yshift=-.5em]h10.east) -- (o5);

\node[inner sep=0.1em,right=of o1,yshift=3em] (rq1) {};
\node[inner sep=0.1em] (rq2) at ([yshift=-1em]rq1){};

\node[inner sep=0.1em, right=of rq1,xshift=.6em] (ro1) {};
\node[inner sep=0.1em] (ro2) at ([yshift=-1em]ro1){};

\node[draw,rectangle,right=of rq1,yshift=-.5em,xshift=-2.5em] (rh1) {H(4)};

\draw (rq1) -- ([yshift=.5em]rh1.west);
\draw (rq2) -- ([yshift=-.5em]rh1.west);
\draw ([yshift=.5em]rh1.east) -- (ro1);
\draw ([yshift=-.5em]rh1.east) -- (ro2);

\node[right=of rh1,xshift=-1.8em] {:=};

\node[draw,rectangle,below=of rh1.west,
xshift=2.55em,yshift=1em] (ry1) {R$_y(+\pi/4)$};
\node[draw,rectangle,below=of ry1,yshift=2.4em] (ry2) {R$_y(+\pi/4)$};

\node[draw,rectangle,right=of ry1,xshift=-1.5em] (ry3) {R$_y(+\lambda/4)$};
\node[draw,rectangle,right=of ry2,xshift=-1.5em] (ry4) {R$_y(-\lambda/4)$};

\node[draw,rectangle,right=of ry3,xshift=-1.5em] (ry5) {R$_y(-\pi/4)$};
\node[draw,rectangle,right=of ry4,xshift=-1.5em] (ry6) {R$_y(-\pi/4)$};

\draw (ry1.east) -- (ry3.west);
\draw (ry2.east) -- (ry4.west);
\draw (ry3.east) -- (ry5.west);
\draw (ry4.east) -- (ry6.west);

\node[circle,draw,fill=black,inner sep=0.1em] (y1) at ([xshift=.65em]ry1.east) {};
\node[circle,draw,fill=black,inner sep=0.1em] (y2) at ([xshift=.65em]ry2.east) {};

\node[circle,draw,fill=black,inner sep=0.1em] (y3) at ([xshift=.65em]ry3.east) {};
\node[circle,draw,fill=black,inner sep=0.1em] (y4) at ([xshift=.65em]ry4.east) {};

\draw (ry1.west) -- ([xshift=-.3em]ry1.west);
\draw (ry2.west) -- ([xshift=-.3em]ry2.west);

\draw (ry5.east) -- ([xshift=.3em]ry5.east);
\draw (ry6.east) -- ([xshift=.3em]ry6.east);

\draw (y1) -- (y2);
\draw (y3) -- (y4);

\node[draw,rectangle,above=of h1,yshift=-2em, minimum height=3em,xshift=.3em] (h26) {H($2^6$)};

\draw ([yshift=-1em]h26.west) -- ([yshift=-1em,xshift=-.3em]h26.west);
\draw ([yshift=-.6em]h26.west) -- ([yshift=-.6em,xshift=-.3em]h26.west);
\draw ([yshift=-.2em]h26.west) -- ([yshift=-.2em,xshift=-.3em]h26.west);
\draw ([yshift=.2em]h26.west) -- ([yshift=.2em,xshift=-.3em]h26.west);
\draw ([yshift=.6em]h26.west) -- ([yshift=.6em,xshift=-.3em]h26.west);
\draw ([yshift=1em]h26.west) -- ([yshift=1em,xshift=-.3em]h26.west);

\draw ([yshift=-1em]h26.east) -- ([yshift=-1em,xshift=.3em]h26.east);
\draw ([yshift=-.6em]h26.east) -- ([yshift=-.6em,xshift=.3em]h26.east);
\draw ([yshift=-.2em]h26.east) -- ([yshift=-.2em,xshift=.3em]h26.east);
\draw ([yshift=.2em]h26.east) -- ([yshift=.2em,xshift=.3em]h26.east);
\draw ([yshift=.6em]h26.east) -- ([yshift=.6em,xshift=.3em]h26.east);
\draw ([yshift=1em]h26.east) -- ([yshift=1em,xshift=.3em]h26.east);

\node[right=of h26,xshift=-1.8em] {:=};

\end{tikzpicture}
\caption{The H($2^6$) gate fabric from~\cite{Anselmetti_2021}.}
\label{fig:h4}
\end{figure}

\paragraph*{Local Properties and Over-Approximation}
For systems with many qubits, it will  be necessary to find strategies for encoding only the part of the system that pertains to a particular proof
and for over-approximating expressions that are 
hard to analyze. 
We present these ideas on a second example: 
the gate fabric on the left of Figure~\ref{fig:h4} preserves the hamming weight of the state
and we want to check this.
Let $\mathrm{H64}[\psi_1,\psi_2]$ be
the set of expressions that describe 
the effect of H($2^6$) on input
state $\psi_1$ and output state $\psi_2$
and
let 
$\mathrm{H4^{i}}[\psi_1,\psi_2]$ be 
expressions that encode the effect
of the $H(4)$ circuit on the $i$-th and $(i+1)$-th
qubits between $\psi_1$ and $\psi_2$.  
Finally, let $\mathrm{HW}(\psi)$ be the arithmetic
expression for the expected Hamming weight
of $\psi$. 
To show that H$(2^6)$ preserves the Hamming 
weight, we have to check that 
$\mathrm{H64}[\psi_1,\psi_2] ~\models ~
 \mathrm{HW}[\psi_1] = \mathrm{HW}[\psi_2]$.
%
%
%
We can decompose this proof into a series of 
smaller proofs that show the preservation of
the expected Hamming weight for the individual
H(4) gates but we can also apply two 
more optimizations.

\smallskip
\noindent
\textit{Over-Approximation.}
Using the matrix representation of H(4) shown 
in Figure~\ref{fig:h4} and a relaxation on
$s$ and $c$, namely only requiring that 
$0\leq s,c \leq 1$ and that $s^2+c^2 = 1$,
we can further simplify the proof obligations
and remove trigonometric functions, making it 
possible to encode the checks in a logic 
that is suitable for of-the-shelf SMT solvers.
In this example, we know from the literature
that the matrix describes the effect of H(4).
An automated approach will have proof
the correctness of abstractions --- similar to
loop invariants in the verification of
classic programs.


\smallskip
\noindent
\textit{Local Properties.}
Finally, we can reduce the proof obligations much more drastically by making an observation that is outside of the scope of automated analysis: a two-qubit gate that preserves the expected Hamming weight of a two-qubit 
state will also preserve the expected Hamming weight of a larger system when applied to two qubits of that system.
Let $G$ be some two-qubit gate on qubits $i$ and $j$ that preserves the expected Hamming weight on a two-qubit system.
%
Since $G$ only affects qubits $i$ and $j$, applying $G$ does not alter the total expected Hamming weight of the $n$-qubit state\footnote{In the $n$-qubit system, the total expected Hamming weight is:
     $\sum_{k \neq i,j} \langle \psi | \frac{1 - Z_k}{2} | \psi \rangle + \langle \psi | \frac{1 - Z_i}{2} + \frac{1 - Z_j}{2} | \psi \rangle$}.
This observation reduces our proof obligation to the proof 
that the $H(4)$ block, given by its matrix 
representation in Figure~\ref{fig:h4}, preserves the 
expected Hamming weight on a two-qubit circuit.
In general, however, it will not be this simple to reduce proof obligations or obvious how to do it. We conjecture
that static analysis and data-flow analysis
can be used for finding beneficial decompositions. 

\section{Demonstration}

We demonstrate the feasibility and efficiency of
compositional verification of quantum circuits
for the two examples presented in the 
previous sections. 
We have generated different verification conditions 
in the \texttt{SMTLib} format: for the H-CNOT 
circuit, we generate a monolithic encoding as well
as the compositional proofs from 
Figure~\ref{fig:decomposed}. The [H+CNOT, P+A1] proof
doubles as the proof obligation in a weakest
precondition approach.
For the H($2^6$) circuit, we have generated 
precise encoding using trigonometric functions as 
well as over-approximating encoding that use the 
matrix that summarizes the effect of H(4) components.
To analyze scalability, we generate variants that 
check preservation of Hamming weights only 
for the first $k$ H(4) components.
Finally, we have generated the 
verification condition for hamming weight 
preservation of H(4) in a two-qubit system.

\begin{table}[t!]
\centering
\footnotesize
\caption{Experimental results for verification of examples from previous sections. 
$\dagger$: for approximated $1/\sqrt{2}$, 
DNS: did not attempt to solve,
DNF: timeout after 30 min.}
\label{tbl:compilation}
\begin{tabular}{|l|r|r|r|r|r|r|r|} \hline
Example & \multicolumn{3}{|c|}{Encoding}  & \multicolumn{2}{|c|}{Analysis} \\
        & Vars & Ass. & Logic &  
        Res. & wct~[s] \\
\hline
H+CNOT           &      25 &     26 & LRA$^\dagger$ & $\checkmark$ & 0.005 \\
H+CNOT, C1        &      17 &     11 & LRA$^\dagger$ & $\checkmark$ &  0.005 \\
H+CNOT, C2        &      17 &     11 & LRA$^\dagger$ & $\checkmark$ & 0.003 \\ \hline

H+CNOT, P+A1      &       9 &      3 & LRA$^\dagger$ & $\checkmark$ & 0.004 \\ \hline
H($2^6$), naive   & 10\,370 & 5\,191 & TRIG & -            & DNS \\
H($2^6$), precise &  3\,330 & 1\,671 & TRIG & -            & DNS \\
H($2^6$)          &  1\,412 &    647 & NRA  & d/k          & DNF \\
H($2^6$), 9/10    &  1\,284 &    583 & NRA  & $\checkmark$ & 8.25 \\
H($2^6$), 8/10    &  1\,156 &    519 & NRA  & $\checkmark$ & 2.29 \\
H($2^6$), 7/10    &  1\,028 &.   455 & NRA  & $\checkmark$ & 1.59 \\
H($2^6$), 5/10    &     772 &.   327 & NRA  & $\checkmark$ & 0.23 \\
H($2^6$), 1/10    &     260 &     71 & NRA  & $\checkmark$ & 0.02 \\ \hline
H(4)              &      20 &     15 & NRA  & $\checkmark$ & 0.01 \\
\hline
\end{tabular}
\end{table}

We have checked all generated SMT problems with 
the Z3 constraint solver, version 4.13.3. 
Experiments were executed on an 
Apple M3 Pro MacBook with 36 GB of RAM, 
running macOS 14.6. We have repeated all 
experiments five times and report results
in Table~\ref{tbl:compilation}. The table
shows number of variables and assertions 
for every verification condition as well 
as the used logic, obtained verdict, 
and average wall-clock runtime.

The experiments show that the proposed techniques
(decomposition, over-approximation, and computing
weakest pre-conditions) are effective:
for the H-CNOT circuit, checking the weakest 
pre-condition requires less than half the 
variables and significantly fewer assertions
than the other proofs.
For the H($2^6$) circuit, monolithic verification
times out at 30 minutes, while analyzing one H(4) 
component finishes in 20 ms. Checking the 
Hamming weight preservation of H(4) on a two-qubit
system is even more efficient.

\section{Conclusion}

We have presented some ideas for scaling 
logic-based verification of quantum circuits
through decomposition and abstraction.
We have demonstrated the effect of these techniques
on two small examples.
The presented ideas can hopefully provide a basis for 
scaling verification for quantum 
circuits with an SMT solver, especially for 
circuits where local properties and clusters 
of gates can be isolated.


\bibliography{main}{}

\end{document}